\documentstyle[twocolumn,aps,prl,floats,epsf]{revtex}

\begin{document}

\wideabs{

\title{Search for narrow $\bar p p$ resonances in the reaction $\bar p
       p \to \bar p p \pi^+ \pi^-$}

\author{
The JETSET Collaboration\\
A.~Buzzo$^5$,
P.~Debevec$^6$,
D.~Drijard$^2$,
R.A.~Eisenstein$^6$,
C.~Evangelista$^1$,
W.~Eyrich$^3$,
H.~Fischer$^4$,
J.~Franz$^4$,
R.~Geyer$^3$,
N.H.~Hamann$^2$\cite{byline},
P.G.~Harris$^6$,
D.W.~Hertzog$^6$,
S.A.~Hughes$^6$,
T.~Johansson$^9$,
R.T.~Jones$^2$,
K.~Kilian$^7$,
K.~Kirsebom$^{5}$,
H.~Korsmo$^8$,
M.~Lo~Vetere$^5$,
M.~Macri$^5$,
M.~Marinelli$^5$,
M.~Moosburger$^3$,
B.~Mou\"ellic$^2$,
W.~Oelert$^7$,
S.~Ohlsson$^{2}$,
A.~Palano$^1$,
S.~Passaggio$^5$,
J.--M.~Perreau$^2$,
M.G.~Pia$^5$,
S.~Pomp$^4$,
M.~Price$^2$,
P.E.~Reimer$^6$,
J.~Ritter$^6$,
E.~Robutti$^5$,
K.~R\"ohrich$^7$,
M.~Rook$^7$,
E.~R\"ossle$^4$,
A.~Santroni$^5$,
H.~Schmitt$^4$,
T.~Sefzick$^7$,
O.~Steinkamp$^7$,
F.~Stinzing$^3$,
B.~Stugu$^7$,
and H.~Wirth$^4$ 
}
\address{
$^1$University of Bari and INFN, Bari, Italy\\
$^2$CERN, European Organization for Nuclear Research, Geneva, Switzerland\\
$^3$University of Erlangen--N\"urnberg, Erlangen, Germany\\
$^4$University of Freiburg, Freiburg, Germany\\
$^5$University of Genova and INFN, Genova, Italy\\
$^6$University of Illinois at Urbana--Champaign, Urbana, Illinois, U.S.A.\\
$^7$Institut f\"ur Kernphysik, Forschungszentrum J\"ulich, J\"ulich, Germany\\
$^8$University of Oslo, Oslo, Norway\\
$^9$Uppsala University, Uppsala, Sweden
}
\date{22 April 1997}

\maketitle

\begin{abstract}
The reaction $\bar p p \to \bar p p \pi^+ \pi^-$ has been studied with
high statistics at CERN-LEAR with incident $\bar p$ momenta from 1.65
to 2.0 GeV/$c$ by the JETSET (PS202) experiment. The aim of this paper
is to search for narrow resonances decaying to $\bar p p$.  No
evidence for such structures is found. In particular, an upper limit
for the production of a 2.02 GeV state with a width of $\Gamma$ = 20
MeV, having been seen in other hadroproduction experiments, is
established.  Our results restrict the cross section for such a peak
to be below $200$ nb at the 95\% confidence level.
\end{abstract}
} %ending the \wideabs

The search for narrow ($\Gamma \approx 10$ MeV) baryonium resonances
decaying to $N \bar N$, has been an exciting field for several years,
when structures decaying to $\bar p p$ and $\bar p p \pi$ have been
reported in some hadron induced reactions.  In particular, evidence
for two narrow states, at 2.02 and 2.2 GeV, was observed in baryon
exchange $\pi^- p$ interactions at 9 and 12 GeV/$c$ in the reaction
$\pi^- p \to p_f \pi^- (\bar p p)$\footnote{the subscripts $f$ or $s$
indicate a relatively fast or slow particle respectively} at the CERN
$\Omega$ spectrometer~\cite{Benkheiri}.  These states were interpreted
as coming from backward production of baryonium, a suggested $\bar q q
\bar q q$ system of quarks and antiquarks, in association with a fast
forward $N^*$ baryon.

A search for these narrow resonances started soon in several
hadroproduction experiments, in
$\pi^-$~\cite{Chung1}~\cite{Ajaltouni}, $K^-$~\cite{wa60mypap} and
$\bar{p}$~\cite{Chung2} induced reactions, all with negative results.
In particular, these states were not observed in the high-statistics
experiment WA56 at the CERN $\Omega$ spectrometer, which studied the
reactions $\pi^- p \to p_f \pi^- (\bar p p_s)$ at 12 GeV/$c$ and
$\pi^+ p \to p_f \pi^+ (\bar p p_s)$ at 20 GeV/$c$~\cite{Ajaltouni}.
Also, no evidence for such states was seen in central production
experiments which studied the reactions $(\pi^+/p) p \to (\pi^+/p)_f
(\bar p p) p_s$ at 85 GeV/$c$~\cite{wa76mypap1}.
%and at 300 GeV/c~\cite{wa76mypap2}.
Searches for baryonium resonances below the $\bar p p$ threshold have
been conducted at LEAR with negative results~\cite{smith}.  However, a
recent reanalysis of $\bar p d \to p_s 5 \pi$ data~\cite{dalka}
reports evidence for a narrow (10 MeV) state at 1.870 GeV.

A similar history took place with another narrow state, at a mass of
2.95 GeV and decaying to $\bar p p \pi^-$, which was reported in the
reaction $\pi^- p \to \bar p p_f \pi^- p_s$, again at the CERN
$\Omega$ spectrometer~\cite{french1}.  Here also a later
high-statistics experiment failed to confirm the existence of such a
state~\cite{french2}.

These experimental results severely dampened the interest in further
searches for baryonium resonances. Recently however the question of
existence of baryonium has been reopened owing to a reanalysis of the
WA56 data. Using a different event selection procedure designed to
enhance the central production of the $\bar p p$ system, the authors
of ref.~\cite{ferrer1} find evidence, in three different reactions,
for a structure at a mass of 2.02 GeV having the narrow width between
10 and 20 MeV.  These data suffer from a relatively high background
due to the fact that all of the analyzed channels are only partially
reconstructed since there is always one missing particle in the final
state.  Nevertheless, the significance of the peaks is quoted to be
between 5 and 7 standard deviations.  This new information motivated
us to search for this narrow resonance in a high statistics experiment
using in-flight $\bar p p$ annihilations.

The main focus of the Jetset (PS202) experiment at CERN-LEAR is the
study of the cross section and spin observables in the reaction $\bar
p p \to \phi \phi \to 4K^{\pm}$.  The motivation for this work is the
search for structures in the observables which are indicative of
hadronic resonances.  The experimental apparatus has been described in
more detail in a previous publication~\cite{jetset}.  The essential
elements which concern the present study are the following.  The
stored $\bar p$ beam interacts with an internal hydrogen gas jet
target in one of the straight sections of LEAR.  The interaction point
is surrounded by a non-magnetic detector which includes
charged-particle tracking chambers, trigger scintillators,
particle-identification detectors (PID), and electromagnetic
calorimetry.  The PID system includes silicon $dE/dx$ pads, freon or
water ``threshold'' \v{C}erenkov counters, and a ring-imaging
\v{C}erenkov counter (RICH)~\cite{jones}, the latter of which was 
introduced after the intial data-taking.  The entire detector is
symmetric in azimuth but is subdivided into a forward ($\approx 10^{0}
< \theta < 45^{0}$) and a barrel ($45^{0} < \theta < 135^{0}$) sector,
where $\theta$ is the laboratory polar angle defined with respect to
the incident antiproton direction.  The detector is open in the
backward direction.  Antiproton momenta in the range from 1.12 to 2.0
GeV/$c$ were utilized.

The trigger was designed to select a sample of four-charged-particle
events with the kinematics commensurate with the reaction $\bar p p
\to 4K^{\pm}$.  This led to a global restriction imposed by the
segmented trigger scintillator system that the emitted particles were
all forward of $\theta = 65^{0}$ and that at least three of them had
$\theta \le 45^{0}$.  A restriction on the $\beta$ of the particles
was imposed such that at least one, and sometimes two, of the
particles would have passed through the threshold \v{C}erenkov
counters without registering a hit.  Events having $\gamma$'s in the
barrel calorimeter were rejected by use of a hardware trigger or by
cuts in the data analysis while those events with $\gamma$'s in the
forward region were identified and removed only in the analysis phase.

One of the background channels, satisfying the trigger conditions and
falling inside the acceptance of the apparatus, is the reaction:
$$\bar p p \to \bar p p \pi^+ \pi^- \eqno(1)$$ This reaction can be
fully reconstructed with little background contamination and it has
been used extensively as a known source of particles for the detector
calibration and the study of systematic errors.  However it is the aim
of this study to make a more detailed analysis of reaction (1) and to
search for narrow resonances decaying to $\bar p p$ in the kinematic
region covered by the experiment.

Since the charged-particle tracks were measured in the absence of a
magnetic field, only the directions are known.  The momenta are thus
computed using the energy-momentum conservation equations given an
assumption on the masses of the four particles.  The three equations
obtained from the 3-momentum conservation can be used to express the
momentum of the particles as linear functions of one parameter, which
we call $\mu$ (the momentum of one of the four outgoing particles).
The sum of the inferred energies of the particles can be expressed in
terms of this parameter as $E_{final}(\mu)$.  Defining
$f(\mu)=E_{final}(\mu) - E_{initial}$ this function has a single
minimum which we call $\Delta E$ and the energy conservation equation
$f(\mu)=0$ has two solutions, of which the ones having a negative
momentum for some of the particles were discarded. The remaining
solutions were then tested for compatibility with the PID system.

Each of the six possible mass combinations in reaction (1) was tested
for compatibility with the kinematics and PID.  If at least one of the
solutions from a given mass combination satisfies these tests, then
the corresponding $\Delta E$ value is plotted.  Distributions of
$\Delta E$ for such solutions at different incident antiproton momenta
are shown in fig. 1.  We observe a dominant peak near $\Delta E = 0$
with little background.  This is a conclusive signature for reaction
(1). A selection of events with $\Delta E \le 40$ MeV defines the
final event sample.
\begin{figure}
  \begin{center}
    \mbox{\epsfxsize=8.6cm\epsffile{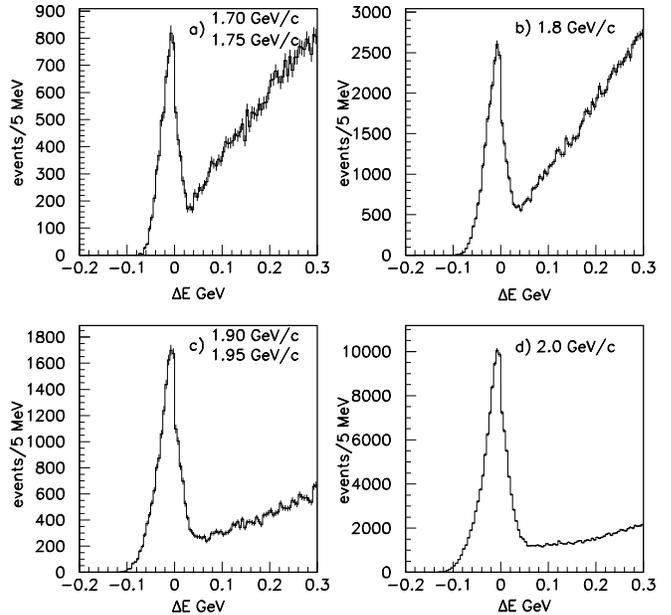}}
  \end{center}

  \caption{$\Delta$E (see text) distributions for reaction $\bar p p
  \to \bar p p \pi^+ \pi^-$ for different $\bar p$ incident momenta.}

  \label{figone}
\end{figure}

In order to derive the cross section for this reaction, the $\Delta E$
distributions were fit using a second order polynomial to describe the
background, and a peak shape derived from Monte-Carlo simulations.
The Monte Carlo is based upon GEANT~\cite{geant} and contains a
detailed description of the apparatus.  Events so produced were
subjected to the same reconstruction procedure and selection criteria
used for the real data.  The acceptance of the apparatus for reaction
(1) was determined from this study.

Due to the limitations imposed mainly by the trigger, the acceptance
function was limited and therefore information on the dynamics of the
reaction cannot be gained from our data only.  We therefore used in
the Monte-Carlo simulation a model which assumed that reaction (1)
proceeds entirely through $\bar p p \to \Delta^{--} \Delta^{++}$ and a
four momentum $t$-distribution from the incident $\bar p$ to the
$\Delta^{--}$ obtained from the study of the same reaction at 3.2
GeV/$c$~\cite{pppipi}. Uncertainties on the validity of this
assumption introduce a systematic error in the determination of the
cross section of approximately 5\%.

One background reaction which survives the selection criteria used to
isolate reaction (1) comes from the process: $$\bar p p \to \Lambda
\bar \Lambda \to (\bar p \pi^+) (p \pi^-) \eqno(2)$$ where the two
$\Lambda$'s have a decay close to the interaction vertex and are
therefore merged into a four pronged event.  The cross section for
this reaction is well known~\cite{llbar} and a Monte-Carlo simulation
of reaction (2) has been performed in order to obtain the overall
acceptance for this process. The estimated contamination is then
subtracted from the data in the computation of the final cross section
for reaction (1).
%The yield of the correction is also shown in table 1.

We obtained, finally, the cross section for reaction (1) which is
listed in table 1 along with the integrated luminosities, and the
estimated contamination from reaction (2).  The data are also shown in
fig. 2 where a comparison with previous measurements~\cite{lys}
~\cite{estman}is given.  The errors on the final cross section values
which are listed in the table are statistical only.  Inspection of the
two distinct runs at 2.0 GeV/$c$ where the final cross sections differ
by approximately $25\%$ in two data sets having different detector and
trigger conditions is indicative of the largest systematic uncertainty
which remains in the scale of the final cross sections. We conclude,
therefore that an overall systematic uncertainty of $25\%$ should be
applied to all the cross sections measured in the present experiment.
\begin{table}

  \caption {Summary results for the study of the reaction $\bar p p
  \to \bar p p \pi^+ \pi^-$.  The factor $N_{\bar \Lambda \Lambda}$
  represents the estimated number of $\bar pp \pi^+ \pi^-$ events
  coming from the reaction $\bar p p \to \bar \Lambda \Lambda$.  The
  errors on the final cross section are statistical only; they do not
  reflect the uncertainties in acceptance owing to different trigger
  and detector configurations.}

  \label{tab:exo}
\begin{tabular}{cccccc}
p  & Lum. & Acceptance & & & $\sigma$\\  
(GeV/$c$) & ($nb^{-1}$) & (\%$_0$) & $N_{\bar pp \pi^+ \pi^-}$ &
 $N_{\bar \Lambda \Lambda}$ & ($\mu$b) \\ \hline  
1.65 & 13.7 & 2.5 & 934 & 278 & $18.9 \pm 1.9$ \\
1.70 & 17.9 & 2.8 & 2559 & 471 & $41.8 \pm 2.1$\\
1.75 & 14.5 & 3.2 & 3212 & 448 & $59.3 \pm 2.2$ \\ 
1.80 & 22.3 & 3.6 & 11049 & 800 & $128.7 \pm 2.3$\\
1.90 & 7.8 & 4.7 &8003 &312 & $211.6 \pm 3.6$\\ 
1.95 & 7.7 & 5.3 & 7843 & 323 & $184.6 \pm 3.4$\\ 
2.0 & 14.5 & 6.4 & 24417& 676 & $ 255.8 \pm 2.1$\\ 
2.0 & 38.5 & 6.0 & 78542 &1795 & $331.1 \pm 1.7$\\
\end{tabular}
\end{table}
\begin{figure}
  \begin{center}
    \mbox{\epsfxsize=8.6cm\epsffile{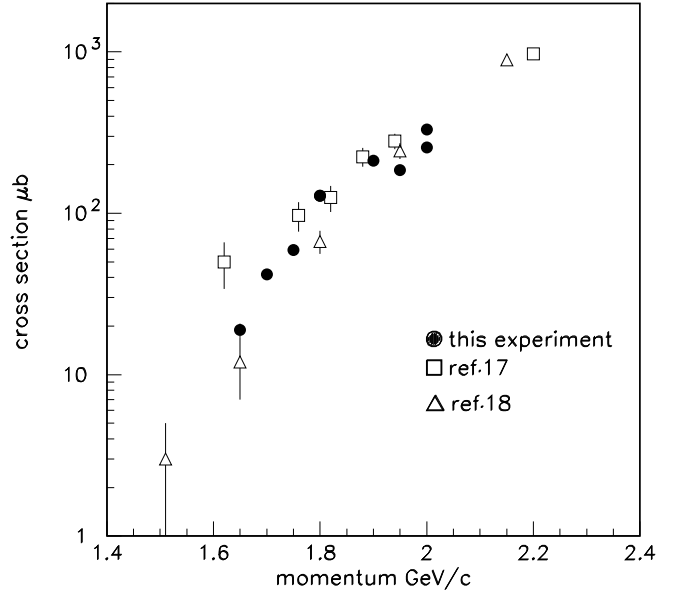}}
  \end{center}

  \caption{The total cross section for the reaction $\bar p p \to \bar
  p p \pi^+ \pi^-$ as measured in this experiment (solid circles) and
  in previous measurements from \protect\cite{lys} (open squares)
  and \protect\cite{estman} (open triangles).}

  \label{figtwo}
\end{figure}

\begin{figure}
  \begin{center}
    \mbox{\epsfxsize=8.6cm\epsffile{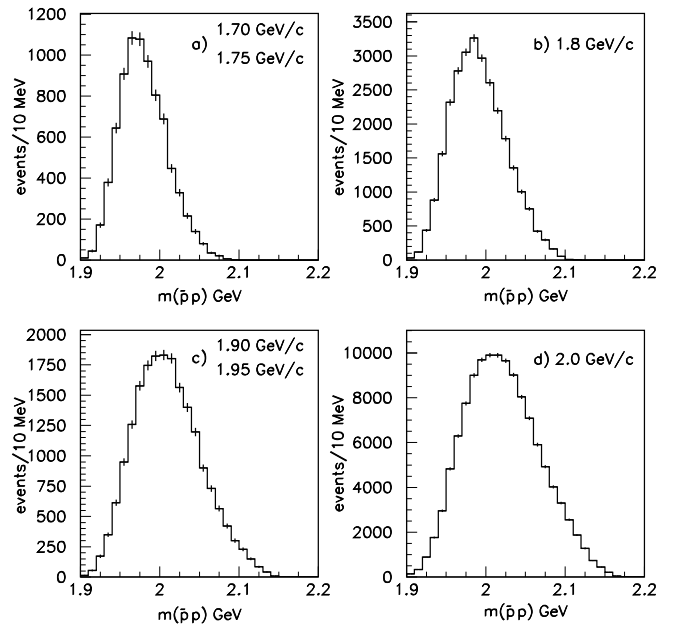}}
  \end{center}

  \caption{$\bar p p$ effective mass distributions derived from events
  from the reaction $\bar p p \to \bar p p \pi^+ \pi^-$ for different
  incident antiproton momenta.}

  \label{figthree}
\end{figure}

The main aim of this study is the analysis of the $\bar p p$ effective
mass distributions for which a precise knowledge of the total cross
section is not important.  The $\bar p p$ mass experimental
resolution, $\sigma_{m(\bar p p)}$, has been computed using the
Monte-Carlo simulation.  At an invariant mass of 2 GeV,
$\sigma_{m(\bar p p)}$~=~9 MeV.  The resulting effective mass
distributions, binned in 10 MeV increments, are shown in fig. 3 for
four groupings of the incident antiproton momenta.  All of the
distributions are smooth and none show striking evidence for resonant
structures.

Low energy $\bar p p$ annihilations cannot be described by peripheral
models, therefore ``central production'' of a given system has no
meaning.  In order to enhance possible contributions from a kinematic
region where the $\bar p p$ system is ``central'', we have combined
the distributions (fig. 4a) and have plotted (fig. 4b) the Feynman
$x_F$ distribution ($x_F = p^*_L/p^*_{Lmax}$) for all of the
data. Here $p^*_L$ is the longitudinal centre of mass momentum of the
$\bar p p$ system with respect to the beam direction. Requiring $|x_F|
\le 0.1$ yields the $m_{\bar p p}$ distribution shown in fig. 4c
where, again, no structure is visible.

The $\bar p p$ effective mass distributions were then fitted using a
smooth shape of the form $(m - m_{th})^{a + b m} e^{-cm - d m^2}$
where a, b, c, d and $m_{th}$ are free parameters.  The residuals
distribution for different $\bar p$ incident momenta are shown in
fig. 4d) and do not show evidence for resonance structure in the 2.02
GeV region.
\begin{figure}
  \begin{center}
    \mbox{\epsfxsize=8.6cm\epsffile{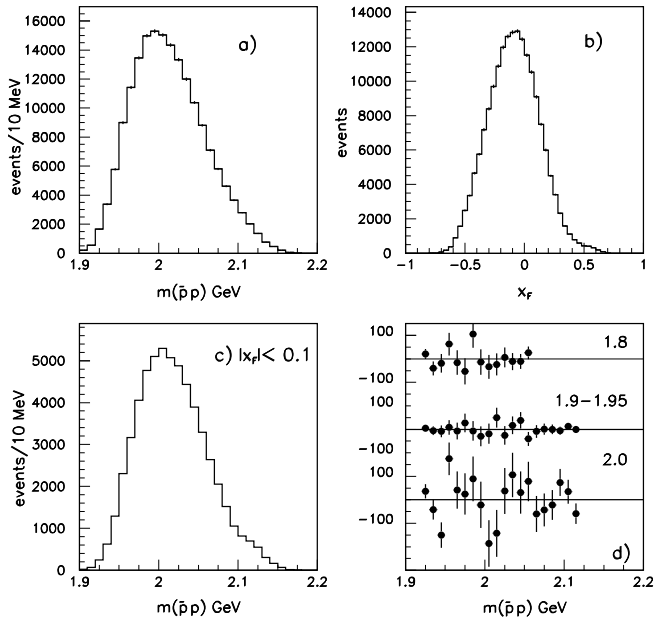}}
  \end{center}

  \caption{a) $m_{\bar p p}$ distribution for all of the data; b)
  Feynman $x_F$ distribution for the $\bar p p$ system; c) $m_{\bar p
  p}$ distribution when $|x_F| \le 0.1$; d) Residuals from the fits
  for different $\bar p$ incident momenta.}

  \label{figfour}

\end{figure}
Attempts to include a Gaussian function describing the presence of a
possible narrow resonant state in the 2.02 GeV mass region having a
width between 10 and 20 MeV failed.  An upper limit for the production
of a 20 MeV wide resonance at 2.02 GeV is determined to be 
\[\sigma < 200 \text{ nb at } 95\% \text{ c.l.} \]

In conclusion, we have studied the reaction $\bar p p \to \bar p p
\pi^+ \pi^-$ with the Jetset experiment at LEAR for a variety of
incident antiproton momenta from 1.65 to 2.0 GeV/$c$.  In addition to
a determination of the cross section, a search was made in the $\bar p
p$ effective mass distribution for possible resonant behavior.  Even
with a selection of events which enhanced the centrality of the $\bar
p p$ system, no structure was observed.  In particular, an upper limit
of $\sigma < 200$ nb at 95\% c.l.  is established for the production
of a 20 MeV wide resonance at 2.02 GeV decaying to $\bar p p$.  Such a
state has been reported to be centrally produced by baryon exchange.
The reaction studied in this paper, in the "central" selection
described above, just selects a baryon exchange mechanism, but no
signal is observed.

This work has been supported in part by CERN, the German
Bundesminister fuer Bildung, Wissenschaft, Forschung und Technologie
(BMBF), the Italian Istituto Nazionale di Fisica Nucleare, the Swedish
Natural Science Research Council, the Norwegian Natural Science
Research Council, and the United States National Science Foundation,
under contract NSF PHY 94-20787.
 
We acknowledge useful discussions with A. A. Grigoryan.


\begin{references}

\bibitem[\dag]{byline}deceased.

\bibitem{Benkheiri}
P. Benkheiri {\em et al.}, Phys. Lett. {\bf  68B} (1977) 483.

\bibitem{Chung1}
S.U. Chung {\em et al.}, Phys. Rev. Lett. {\bf  45} (1980) 1611.

\bibitem{Ajaltouni}
Z. Ajaltouni {\em et al.}, Nucl. Phys. {\bf B209} (1982) 301.

\bibitem{wa60mypap}
T. Armstrong {\em et al.}, Nucl Phys. {\bf B206} (1982) 185.

\bibitem{Chung2}
S.U. Chung {\em et al.}, Phys. Rev. Lett. {\bf 46} (1981) 395.

\bibitem{wa76mypap1}
T. Armstrong {\em et al.}, Z. Phys. C {\bf 35} (1987) 167.

%\bibitem{wa76mypap2}
%T. Armstrong {\em et al.}, CERN/EP 88-124, 23 september 1988.

\bibitem{smith}
A. Angelopoulos {\em et al.}, Phys. Lett. {\bf 159B} (1985) 210.

\bibitem{dalka}
O.D. Dalkarov {\em et al.}, Phys. Lett. {\bf 392B} (1997) 229.

\bibitem{french1}
C. Evangelista {\em et al.}, Phys. Lett. {\bf 72B} (1977) 139.

\bibitem{french2}
T. Armstrong {\em et al.}, Phys. Lett. {\bf 85B} (1979) 304.

\bibitem{ferrer1}
A. Ferrer {\em et al.}, Nucl. Phys. {\bf A558} (1993) 191c and
A. Ferrer {\em et al.}, Proceedings of the Hadron95 Conference,
Manchester, July 1995.
%\bibitem{ferrer2} (we don't need two references for this stuff.)

\bibitem{jetset}
L. Bertolotto {\em et al.}, Phys. Lett. {\bf B345} (1995) 325.

\bibitem{jones}
R.T. Jones {\em et al.}, Nucl. Instr. and Meth. {\bf A343} (1994) 208
and S. Passaggio {\em et al.}, Nucl. Instr. and Meth. {\bf A371}
(1996) 188.

\bibitem{geant}
R. Brun {\em et al.}, GEANT3, CERN Report DD/EE/84-1 (1987).

\bibitem{pppipi}
Y. Oren {\em et al.}, Nucl. Phys. {\bf B53} (1973) 269.

\bibitem{llbar}
P.D. Barnes {\em et al.}, Nucl. Phys. {\bf A526} (1991) 575.

\bibitem{lys}
J. Lys {\em et al.}, Phys. Rev. {\bf D7} (1973) 610.

\bibitem{estman}
P.S. Eastman {\em et al.}, Nucl. Phys. {\bf B51} (1973) 29.

\end{references}
\end{document}